\documentclass[a4paper,10pt]{article}
\usepackage[utf8]{inputenc}
\usepackage{times,psfrag,rotating}
\usepackage{color}
\usepackage{amsmath,graphics,float,graphicx,caption,subcaption,cite}
\usepackage{jcappub}

\usepackage{soul}

\usepackage{graphicx}
\usepackage{ulem}

\def\xb{\bar{x}_{{\rm H{\sc I}}}\,}
\def\HI{H{\sc i}\,}

\def\x{{\bf \textit{x}}}

\title{The effects of the small-scale DM power  on the cosmological neutral hydrogen (\HI) distribution at high redshifts}

\author[1,2]{Abir Sarkar,}
\author[3,4]{Rajesh Mondal,}
\author[5]{Subinoy Das,}
\author[1]{Shiv.K.Sethi,}
\author[3,4]{Somnath Bharadwaj,}
\author[6]{David J. E. Marsh}

\affiliation[1]{Raman Research Institute,\\ Bangalore, India}
\affiliation[2]{Indian Institute of Science,\\ Bangalore, India}
\affiliation[3]{Department of Physics, Indian Institute of Technology Kharagpur,\\ Kharagpur - 721302, India}
\affiliation[4]{Centre for Theoretical Studies, Indian Institute of Technology Kharagpur,\\ Kharagpur - 721302, India}
\affiliation[5]{Indian Institute of Astrophysics,\\ Bangalore,India}
\affiliation[6]{Department of Physics, King’s College London,\\ Strand, London, WC2R 2LS, United Kingdom }

\emailAdd{abir@rri.res.in}
\emailAdd{rm@phy.iitkgp.ernet.in}

\abstract{The particle nature of dark matter remains a mystery. In this 
paper, we consider two dark matter models---Late Forming Dark Matter (LFDM) and Ultra-Light Axion (ULA) models---where the matter
power spectra show novel effects on small scales.  The  high redshift
universe offers a powerful probe of their parameters.  In particular, we study 
two cosmological observables: the neutral hydrogen (HI) redshifted 21-cm signal 
from the epoch of reionization, and the evolution of the collapsed 
fraction of HI in the redshift range $2 < z < 5$.  We model the theoretical predictions of the models
using CDM-like N-body simulations with modified initial conditions, and generate reionization fields using
an excursion-set model. The N-body approximation is valid on the length and halo mass scales studied.
 We show that LFDM and 
ULA models predict an increase in the HI power spectrum from the epoch of 
reionization by a factor between 2--10 for a range of scales $0.1<k<4 \, \rm Mpc^{-1}$. Assuming a fiducial model where a neutral hydrogen fraction $\bar{x}_{HI}=0.5$ must be achieved by $z=8$, the reionization process allows us to put approximate bounds 
on the redshift of dark matter formation $z_f > 4 \times 10^5$ (for LFDM) and the axion mass $m_a > 2.6 \times 10^{-23} \, \rm eV$ (for ULA). The comparison of the collapsed mass
fraction inferred from damped Lyman-$\alpha$ observations to the theoretical predictions of  our models lead to the weaker bounds: $z_f >  2 \times 10^5$
and $m_a > 10^{-23} \, \rm eV$. These bounds are consistent with other constraints in the literature using different observables; we briefly discuss
how these bounds compare with possible constraints from the observation of 
luminosity function of galaxies at high redshifts.  In the case of ULAs,
these constraints  are also consistent with a solution to the cusp-core problem of CDM.

}

\arxivnumber{1512.03325}

\begin{document}

\maketitle

\section{Introduction}
\label{sec:intro}
After intensive searches over decades, the nature of 
dark matter is still a mystery. This strange type of matter 
reveals  its presence only through gravitational 
interaction in astrophysics and cosmology. The existence of dark matter is very well confirmed 
by various observations  covering many length scales and epochs 
of our universe. Some examples are the cosmic microwave 
background anisotropy 
experiments\cite{Ade:2013zuv,Hinshaw:2012aka,Sievers:2013ica}, 
large scale structure surveys\cite{Tegmark:2006az,Tegmark:2003ud}, 
study of the galaxy rotation curve\cite{Begeman:1991iy}, 
cosmological weak gravitational lensing 
observations\cite{Bartelmann:1999yn} etc. Though all of these experiments 
have confirmed the existence of dark matter, they do not 
throw any light on its fundamental nature. One of the 
leading candidate of dark matter, the Weakly Interacting 
Massive Particle (WIMP) or the traditional cold dark matter (CDM), 
is inspired by the well known WIMP miracle. This WIMP 
miracle relies on the coincidence between weak scale 
cross-section and the dark matter freeze-out cross-section 
needed to produce correct relic density. This discovery 
has driven a lot of direct\cite{Angloher:2011uu,Aprile:2010um,Aprile:2012nq,Ahmed:2010wy,Akerib:2013tjd}, 
indirect\cite{Adriani:2008zr,Adriani:2010rc,Adriani:2008zq,Adriani:2011xv,FermiLAT:2011ab,Ackermann:2010ij,Abdo:2009zk,Barwick:1997ig,Aguilar:2007yf} 
and collider\cite{Goodman:2010yf,Fox:2011pm} searches 
 but these experiments have not yet succeeded  in determining the particle 
nature of dark matter.  In fact, the results from many of these experiments are found 
to be in conflict\cite{Hooper:2013cwa} with each other.

In addition, there exists some long standing cosmological problems 
with WIMPs. One of them is the \textbf{cusp-core}
problem\cite{deBlok:2009sp}, indicated by   the discrepancy between increasing dark matter halo 
profile (cusp) towards the center of galaxy from N-body simulation \cite{Navarro:1995iw} and the observationally 
found relatively flat density profile \cite{Walker:2011zu}. It is instructive to note that though  there exist  difference in opinions about  cusp-core issue of CDM 
and dwarf galaxy observations \cite{Richardson:2014mra}. 
Another issue with WIMPs is the missing satellite 
problem\cite{Klypin:1999uc,Moore:1999nt} where N-body 
simulations of structure formation with CDM produce a 
lot more satellite haloes of a Milky-Way type galaxies than 
observed. The 
``too big too fail''\cite{Garrison-Kimmel:2014vqa,BoylanKolchin:2011de} 
problem, another thorn in the crown of $\Lambda$CDM model, shows that the majority of the most massive 
subhaloes of the Milky Way are too dense to host any of its 
bright satellites. Some recent works claim that  these issues may persist even when the impact of 
small scale baryonic physics is included  \cite{Pawlowski:2015qta,Onorbe:2015ija, 2014arXiv1412.2748S}. All these issues have inspired a drive to go beyond the standard WIMP picture of CDM and consider alternative candidates for dark matter, which differ from CDM on galactic scales (while reproducing its success on cosmological scales.
One such alternative is the warm dark matter (WDM), which causes 
a decrement in  the matter  power at small 
scales\cite{Dolgov:2000ew,Viel:2005qj} due to a larger free-streaming distance than CDM and may provide a solution to some of the small-scale problems 
\cite{0004-637X-556-1-93}. But 
such models are not without issues. The N-body simulation of WDM is very sensitive to WDM mass and it tends to produce fewer
number of satellites than already observed\cite{Polisensky:2010rw,Anderhalden:2012qt,Lovell:2011rd}. 
The status of the cusp-core problem has  
improved\cite{Maccio:2012qf,Schneider:2011yu} but 
unfortunately it is not fully resolved by introducing 
the WDM. In addition WDM has issues with Lyman-$\alpha$ forest flux power spectrum \cite{Viel:2005qj} where observational results, combined with hydrodynamic simulations of structure
formation  place the strongest constraint to date on
WDM mass  $m_W \geq 3.3\, \rm keV$ \cite{Baur:2015jsy}. But the preferred mass range that arises  from WDM N-body simulation is different. It was shown in \cite{Maccio:2012qf} that the solution to
the \textbf{cusp-core} problem requires WDM particles with mass $m_W \sim 0.1\, \rm keV$. 
 It has also been pointed out that WDM particles with mass in the range $ 1.5\, \rm keV  \leq m_W \leq 2\, \rm keV$ 
 can solve the too-big-to-fail problem \cite{Lovell:2011rd}. In this work we investigate two models of dark matter that result in  WDM-like suppression in power at  small scales but  their physical origin and nature of power spectra at  small scale is significantly different from WDM models.\\

We focus on two models of dark matter---namely ``Late Forming Dark Matter (LFDM)" and "Ultra Light Axion (ULA)" dark matter. The former has its origin in extended neutrino physics \cite{Das:2006ht} while the later originates from string theory \cite{Arvanitaki:2009fg}. In Both of these cases,  dark matter starts behaving as CDM after the epoch of BBN and before the epoch of matter radiation equality and both of them  have similar features (oscillations and suppression in small scale power) in matter power spectra. Given the similarities between the ULA and LFDM models, it is reasonable to expect that in some viable region of the ULA
parameter space, the two models may share the same
phenomenology of the matter clustering at small scale. Indeed a recent N-body simulation work on LFDM \cite{Agarwal:2014qca} and two simulations for ULA DM \cite{Schive:2014dra, Marsh:2015wka,Marsh:2015xka} have shown that these models appear to offer a good solution to cusp-core issue while being consistent with large scale clustering.

In this paper, we study the effects of these dark matter models on two distinct features of the 
universe. The first one is the Epoch of Reionization (EoR). 
This epoch is one of the most important milestone of the 
history of the universe, though one of the least well 
understood. From QSO absorption studies and CMB anisotropy measurements
we know  the  reionization occurred at $z \simeq 9$ even though
it could be an extended process lasting until $z \simeq 6.5$ \cite{Ade:2015xua,Fan:2000gq}. The details of EoR are  sensitive to the collapsed 
fraction of baryons which in turn depends on the matter power spectrum
at small scales (see e.g. \cite{Barkana:2000fd}). In particular, we study
the impact of LFDM and ULA models on the HI signal from the EoR. Another 
direct probe of the suppression in power at small scales is the evolution
of collapsed fraction at high redshifts. The   Damped Lyman-$\alpha$ data, 
based on absorption studies,  in the redshift range $2 < z < 5$ can
be used to construct the evolution of the total amount of HI in bound 
objects   \cite{Noterdaeme:2009tp,Peroux:2001ca,Noterdaeme:2012gi,Zafar:2013bha}, which provides a lower limit to the collapsed fraction of matter in this 
redshift range.\\

 To constrain LFDM and ULAs using EoR, one needs the dark matter distribution at high redshift $z=8$.  
We have used a particle-mesh N-body code\cite {Bharadwaj:2004wc, Mondal:2014xma} to evolve the initial matter power spectra of LFDM and ULA DM models from $z=124$ to $z=8$.
This is how WDM simulations are also studied \cite{Polisensky:2010rw,Anderhalden:2012qt,Lovell:2011rd} in the literature and is valid on scales above the present-day Jeans scales of these models.
It is important to check the validity of this procedure for LFDM and ULA DM. This has been done in \cite {Schive:2014dra,Agarwal:2014qca}. It is instructive to note that these simulation will not resolve 
the internal halo structure. For ULAs, this could be achieved using a wave-like simulation following the method of \cite { Schive:2014dra}. But for the purpose this paper, our approach is accurate enough to confront LFDM and ULA models from the Epoch of Reionization (EoR) observations.

The paper is organized as follows. In Section \ref{sec:cosmoLFDM}, we provide a brief description about the particle nature 
of the LFDM and ULAs along with its cosmological effects on the matter 
power spectrum. In section \ref{sec:sim}, we provide the details 
of the simulation set-up for the reionization field. In 
section \ref{sec:result}, we present  the results of our simulations
including the predictions for  the HI power spectrum. In section 
\ref{sec:LFDMonCosmo}, using analytical models for mass function, we  study the effect 
of our models  on the evolution of cosmological gas 
density and compare our results with damped Lyman-$\alpha$  observations. Section \ref{sec:conclude} is reserved for concluding remarks.

Throughout this paper, we have used the Planck+WP best fit values of
cosmological parameters: $\Omega_{\rm m0}=0.3183$,
$\Omega_{\rm \Lambda0}=0.6817$ , $\Omega_{\rm b0}\,h^2=0.02203$, 
$h=0.6704$, $\sigma_8=0.8347$, and $n_{\rm s}=0.9619$ 
\cite{Ade:2015xua,Ade:2013zuv}.

\section{Particle origin and Cosmology of LFDM and ULA DM}
 \label{sec:cosmoLFDM}
\subsection{LFDM}

 Here, we consider a scenario in which the DM is consequence of a phase transition in the dynamics of a scalar
field \cite{Das:2006ht,Sarkar:2014bca}.  There can be two 
types of LFDM: Bosonic and Fermionic. The bosonic or 
scalar LFDM forms when a scalar field $\phi$ (coupled 
with massless neutrino), trapped in the false vacuum 
of potential $V(\phi)$, makes transition to the true 
vacuum during the evolution of the universe and starts 
behaving like dark matter. In case of Fermionic LFDM, 
the freely propagating massless neutrinos get trapped 
into small nuggets by some fifth force \cite{Das:2012kv}
and the nuggets start behaving like dark matter. But 
as both of these two types of LFDM gets their initial 
condition  from  standard model neutrino, we 
don't expect to see any difference in the evolution
of the universe. These models have 
been studied by\cite{Sarkar:2014bca} and it is found 
that the dark matter must form deep inside the radiation 
dominated era. The small scale effects of LFDM is studied
in\cite{Agarwal:2014qca} using $N$-body simulations.

\begin{figure}[H]
\begin{subfigure}{.5\textwidth}
  \centering
  \includegraphics[width=0.75 \linewidth , angle=-90]{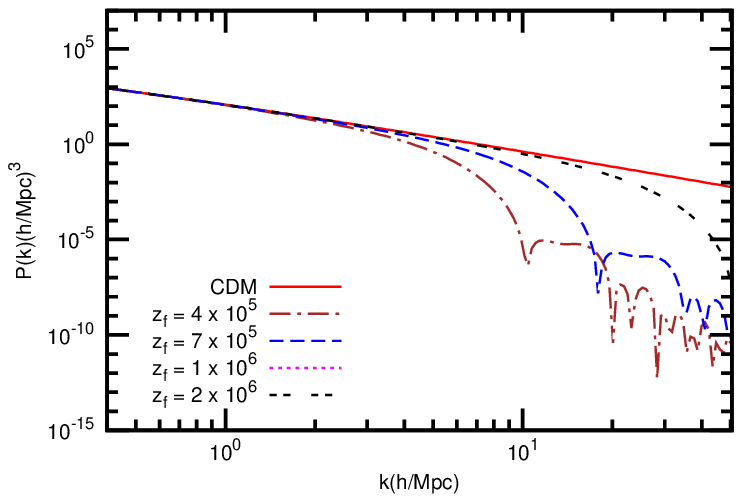}
  \label{fig:1a}
\end{subfigure}%
\begin{subfigure}{.5\textwidth}
  \centering
  \includegraphics[width=0.75 \linewidth , angle=-90]{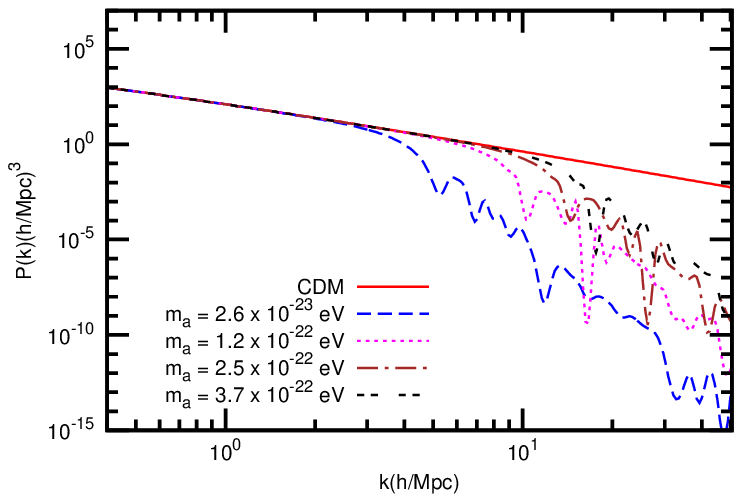}
  \label{fig:1b}
\end{subfigure}
\caption{Normalized power spectra for LFDM models with different $z_f$ (Left panel) and ULA models with different masses (Right Panel)}
\label{fig:1}
\end{figure}

There are two main features of LFDM cosmology 
that manifest themselves in the  matter power spectra. 
First, there is a sharp break in the power at the 
co-moving  scale $k_e = aH_e$; where $H_e$ is the 
Hubble scale  at the epoch of phase transition $z_f$. 
Second, there are  damped oscillations at  smaller 
scales. Both of these can be seen in  Figure~\ref{fig:1}, 
in which we show the normalized power spectra  for LFDM models
for many  four different values of  $z_f$.
The details of the computation of  this power 
spectrum are provided in \cite{Sarkar:2014bca}.

The nature of the power spectrum can be understood 
as follows. Before the phase transition, due 
to strong coupling  with massless 
neutrinos, the density and velocity perturbations 
of the scalar field follow that   of the neutrino. 
On the super horizon scales, the massless neutrinos
behave like other forms of matter such as the CDM. 
However the perturbations in this component are 
washed out owing to free-streaming on scales smaller 
than the  horizon at $z = z_f$. The complete solution for the density 
perturbation is an exponentially damped oscillation 
at sub-horizon scales. The scales smaller than $k_e$, i.e the scale which enters the horizon at the time 
 of transition,  are thus expected to carry the 
signature of the behavior of massless neutrino 
as LFDM obtains its initial conditions from them.
After the transition, the scalar field decouples 
from the neutrino and starts behaving like CDM. It 
can be noted in Figure~\ref{fig:1} that this is 
exactly how LFDM power spectrum behaves for 
$k > k_e$. As $z_f$ is increased the feature 
shifts to larger $k_e$, or smaller scales. As 
$z_f$ tends to infinity, the LFDM matter spectrum 
approaches the $\Lambda$CDM results.

\subsection{ULA dark matter}

  Another candidate for dark matter with novel effects on galactic scales which is also well studied in recent past is the dark matter from ultra light axion like particles (ULAs) 
  \cite{Hu:2000ke, Hlozek:2014lca, Bozek:2014uqa, Amendola:2005ad, Marsh:2010wq, Marsh:2011bf,Schive:2014dra,Marsh:2013ywa,Urena-Lopez:2015gur,Park:2012ru}. For a recent review on 
  this, see Refs.\cite{Marsh:2015xka}. In this case, there are ultra-light axions  with mass as low as $10^{-18} - 10^{-33} eV$ which has their 
 well motivated origin is string axiverse scenario \cite{Arvanitaki:2009fg}. They can play the role of dark matter at late time when the axion field overcome the Hubble friction
 and starts oscillating coherently in a quadratic potential. This dark matter from ULA are similar to the well known QCD axion DM. The only difference being that the 
 mass of these particle is much lower than QCD axion and thus the dark matter behavior happens at much late time when Hubble parameter drops below the mass of ULAs. 
 ULAs with smaller mass start behaving as dark matter later.

ULAs are created by spontaneous symmetry breaking and behave as a coherent classical scalar field. It obtains its initial condition after symmetry breaking in the early universe. Unlike LFDM it remains frozen at this initial value by Hubble drag and behaves like a cosmological constant as long as $H(t) > m_a$ where $m_a$ is the axion mass. When the Hubble parameter drops below $m_a$, the field begins behaving like cold dark matter.  Below a certain scale  determined by a momentum-dependent effective sound
speed of the perturbation in the scalar field, the particle free-streaming  leads to 
 suppression of growth of structures\cite{Hu:2000ke,Marsh:2010wq,Amendola:2005ad,Park:2012ru}. This suppression is indicated as a sharp break in the matter power spectra at the corresponding Jeans scale as shown in Figure~\ref{fig:1}.
 The Jeans scale is given by \cite{Marsh:2015xka} 
 \begin{equation}
  k_J =  66.5 \, a^{\frac{1}{4}}\, \left (\frac{\Omega_a \, h^2}{0.12}\right )^{\frac{1}{4}} \, \left (\frac{m_a}{10^{-22}}\right )^{\frac{1}{2}}\, \rm{Mpc^{-1}}.
\end{equation}
 and after some simplification it can be shown that 
  the suppression of structure formation in the matter dominated era occurs below a scale \cite{Marsh:2010wq}:  
\begin{equation}
  k_m \sim \Bigg(\frac{m}{10^{-33}eV}\Bigg)^{1/3} \Bigg(\frac{100\, \rm{km s^{-1}}}{c}\Bigg)h\, \rm{Mpc^{-1}}.
\end{equation}
This means that the  lesser the axion mass the  larger the suppression scale as can be seen from Figure~\ref{fig:1}. For the least  massive ULA used in this work, the above relation yields $k_m \sim 1\,h 
\rm{Mpc^{-1}}$.  This Jeans scale in linear power spectra  has a considerable effect on CMB physics and  ULA can also be constrained from large scale structure surveys.
The cosmologically relevant mass range for ultra light axion is given by $ 10^{-33} \rm eV < m_a < 10^{-18} \rm eV $ \cite{Marsh:2015xka}. recently, using Planck and WiggleZ data and linear physics alone Hlozek et al \cite{Hlozek:2014lca} essentially excludes mass range below $m_a < 10^{-24}$ if ULA constitutes 
the  whole of  dark matter.  In this work, we are interested in ULAs with masses $\gtrsim 10^{-23}$~eV which suppress structure formation on scales comparable WDM with $ m_W\gtrsim 0.1 \rm{keV}$

%Amendola and Barbieri \cite{Amendola:2005ad} used 
%the  Lyman-$\alpha$ data to constrain the mass of ULAs: $m_a > 5 \times 10^{-23} eV$. \\

\section{Simulating the reionization field}
\label{sec:sim}
The redshift evolution of the mass averaged neutral fraction $\xb$ 
during EoR is largely unknown. It is only constrained  from the CMB anisotropy
and polarization measurements (\cite{Adam:2015rua})  which allow us to infer the    the
optical depth  integrated through the reionization surface. Therefore, the CMB
constraints can be  satisfied for a wide range of ionization histories \cite{Kuhlen:2012vy}.  

Our aim here to 
study the HI signal from the reionization era for a class of LFDM and axion
models and, if possible, discern generic features of such models from the
HI power spectra. Given the uncertainty of reionization history, we 
do not assume a  particular model 
for reionization history $\xb(z)$. Instead  we fix  the redshift and the 
ionization fraction at which these models are compared. We take 
   $z=8$ and  $\xb=0.5$ at this redshift for our simulations.

Our method 
of constructing reionization fields consists of three steps: 
(i) generating the dark matter distribution at the desired redshift, 
(ii) identifying the location and mass of 
collapsed dark matter halos within the simulation box, 
(iii) generating the neutral hydrogen map using an excursion 
set formalism \cite{Furlanetto:2004nh}. The assumption here is 
that the hydrogen exactly traces the dark matter field and 
the dark matter halos host the ionizing sources\footnote{The assumption that
the dark matter follows the baryons is justified for simulating the HI signal
from EoR for the following reasons. The issue of simulating and studying this
signal is essentially a two-scale problem: the scale at which the structures
collapse and the scale at which the HI signal is observable. These  scales
are generally separated by orders of magnitude. For instance, the objects
that collapse around $z \simeq 10$  lie in the mass range $M\simeq 10^9\hbox{--}10^{10} \, \rm M_\odot$, which correspond to length scales $L \simeq 0.2\hbox{--}0.4 \, \rm Mpc$ or equivalently $k \simeq 2\pi/L \simeq  30\hbox{--}15 \, \rm Mpc^{-1}$. However, we study the HI signal in the range $0.1 < k < 4 \, \rm Mpc^{-1}$. So even though the density field is highly non-linear at the scale of 
the collapse and therefore the assumption that baryons follow dark matter
is not a good one, it generally is an excellent assumption at the 
scales at which the HI is probed, which lie in the range from mildly non-linear to highly linear at the redshifts of interest.} We discuss 
our method in the following sections.

\subsection{Generating the dark matter density field}
\begin{figure}
\psfrag{pk}[c][c][1][0]{$\Delta^2 (k)$}
\psfrag{k}[c][c][1][0]{$k\,$ (${\rm Mpc}^{-1}$)}
\psfrag{10}[c][c][1][0]{10}
\psfrag{CDM}[c][c][1][0]{{$\Lambda$CDM}\, \,}
\psfrag{LFDM z=2M}[c][c][1][0]{{LFDM $z_f=$2M}\, \, \, \,}
\psfrag{1M}[c][c][1][0]{{1M}}
\psfrag{0.7M}[c][c][1][0]{{0.7M}}
\psfrag{0.4M}[c][c][1][0]{{0.4M}}
\psfrag{ULA mw14}[c][c][1][0]{{$m_a=3.7\times10^{-22}$eV}\, \, \, \, \, \, \, \, \, \,}
\psfrag{mw12}[c][c][1][0]{{$2.5\times10^{-22}$eV}\, \, \, \, \, \, \, \, \,}
\psfrag{mw09}[c][c][1][0]{{$1.2\times10^{-22}$eV}\, \, \, \, \, \, \, \, \,}
\psfrag{mw05}[c][c][1][0]{{$2.6\times10^{-23}$eV}\, \, \, \, \, \, \, \, \,}
\centering
\includegraphics[width=1.05\textwidth]{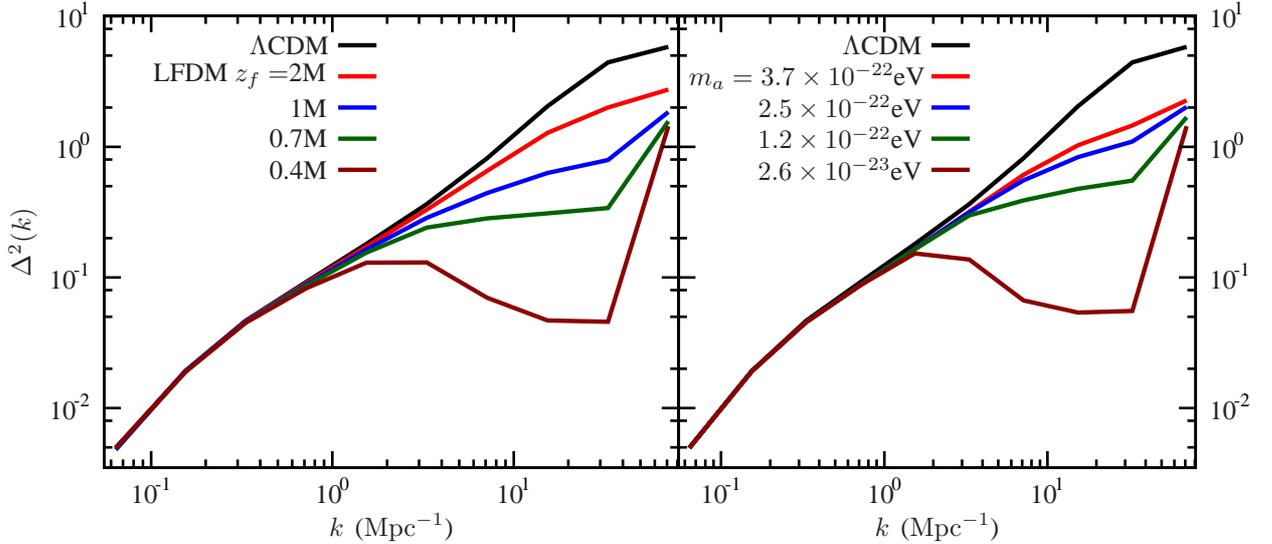}
\caption{This shows the dimensionless matter power spectrum 
$\Delta^2 (k)$ at $z=8$ calculated from the $N$-body 
outputs using four different LFDM models (left panel) with 
$z_f= \{2, 1, 0.7, 0.4\} \times 10^6$ and four different axion DM models 
(right panel) with axion masses $m_a=\{3.7, 2.5, 1.2, 0.26\} \times10^{-22}\,{\rm eV}$. The black 
solid curves are the $\Lambda$CDM power spectrum.}
\label{fig:pk}
\end{figure}

We have used a particle-mesh $N$-body code \cite{Bharadwaj:2004wc} 
to simulate the $z=8$ dark matter distribution. We use the linear power
spectrum  (Figure \ref{fig:1}) to generate the initial Gaussian random density 
field  at  $z=124$. For LFDM, the linear power spectra are
generated using  a modified version of CAMB  \cite{Sarkar:2014bca} and
axionCAMB is  used for the  ULA dark matter \cite{Hlozek:2014lca}. We have
done simulations for the following  models:  
 $\Lambda$CDM, LFDM models with $z_f= \{2, 1, 0.7, 0.4\} \times 10^6$, 
 and ULA models with $m_a=\{3.7, 2.5, 1.2, 0.26\} \times 10^{-22}\,{\rm eV}$. 
Figures~\ref{fig:pk} shows the dimensionless matter 
power spectrum $\Delta^2 (k)=k^3 P (k)/2 {\rm \pi}^2$  output from the N-body
simulation at $z=8$. Note the suppression of power on  scales $k \gtrsim
1\,{\rm Mpc}^{-1}$  for both the  LFDM (left panel) and ULA (right panel)
models compared to the $\Lambda$CDM model. This suppression 
deepens as we  decrease the value of $z_f$ and $m_a$ in the LFDM  and ULA
models respectively. 

Our  simulation volume is a $150\, {\rm Mpc}^3$ comoving 
box. We have run our simulation with a $4288^3$
grid using  $2144^3$ dark matter particle. The spatial resolution is 
$35\, {\rm kpc}$ which corresponds to a mass resolution of 
$1.36 \times10^7\, {\rm M_{\odot}}$. The $N$-body code used for this work has been 
parallelized \cite{Mondal:2015oga} for shared-memory machines 
using Open{\bf MP}.

\subsection{Identifying collapsed dark matter halos}
We have used a friends-of-friends (FoF) halo finder code \cite{Mondal:2014xma} 
to identify the location and mass of the collapsed dark matter 
halos from the outputs of the $N$-body simulation. We have used a 
fixed linking length $0.2$ times the mean inter-particle separation 
and require a halo to have at least $10$ dark matter particles 
which corresponds to a minimum halo mass of 
$1.36 \times10^8\, {\rm M_{\odot}}$. Our 
choice of the minimum halo mass is well motivated  because a  halo mass 
of a few $10^8\,{\rm M_{\odot}}$ \cite{Mesinger:2007pd} (at 
$z=8$)  also corresponds to the  virial temperature 
($\sim 10^4$K) of HI-cooling threshold. We do not consider the possible
impact of  $H_2$-cooled
mini-haloes, which lie in the mass range from Jeans' mass to $10^8 \, \rm M_\odot$, on reionization.   

\begin{figure}
\psfrag{dndlnm}[c][c][1][0]{$\frac{{\rm d}n}{{\rm d(ln}M)}$ ($h^3 {\rm Mpc}^{-3}$)}
\psfrag{m}[c][c][1][0]{$M$ ($10^{10} \, h^{-1}\, {\rm M_{\odot}}$)}
\psfrag{10}[c][c][1][0]{10}
\psfrag{CDM}[c][c][1][0]{{$\Lambda$CDM}\, \,}
\psfrag{LFDM z=2M}[c][c][1][0]{{LFDM $z_f=$2M}\, \, \, \,}
\psfrag{1M}[c][c][1][0]{{1M}}
\psfrag{0.7M}[c][c][1][0]{{0.7M}}
\psfrag{0.4M}[c][c][1][0]{{0.4M}}
\psfrag{ULA mw14}[c][c][1][0]{{$m_a=3.7\times10^{-22}$eV}\, \, \, \, \, \, \, \, \, \,}
\psfrag{mw12}[c][c][1][0]{{$2.5\times10^{-22}$eV}\, \, \, \, \, \, \, \, \,}
\psfrag{mw09}[c][c][1][0]{{$1.2\times10^{-22}$eV}\, \, \, \, \, \, \, \, \,}
\psfrag{mw05}[c][c][1][0]{{$2.6\times10^{-23}$eV}\, \, \, \, \, \, \, \, \,}
\centering
\includegraphics[width=1.05\textwidth]{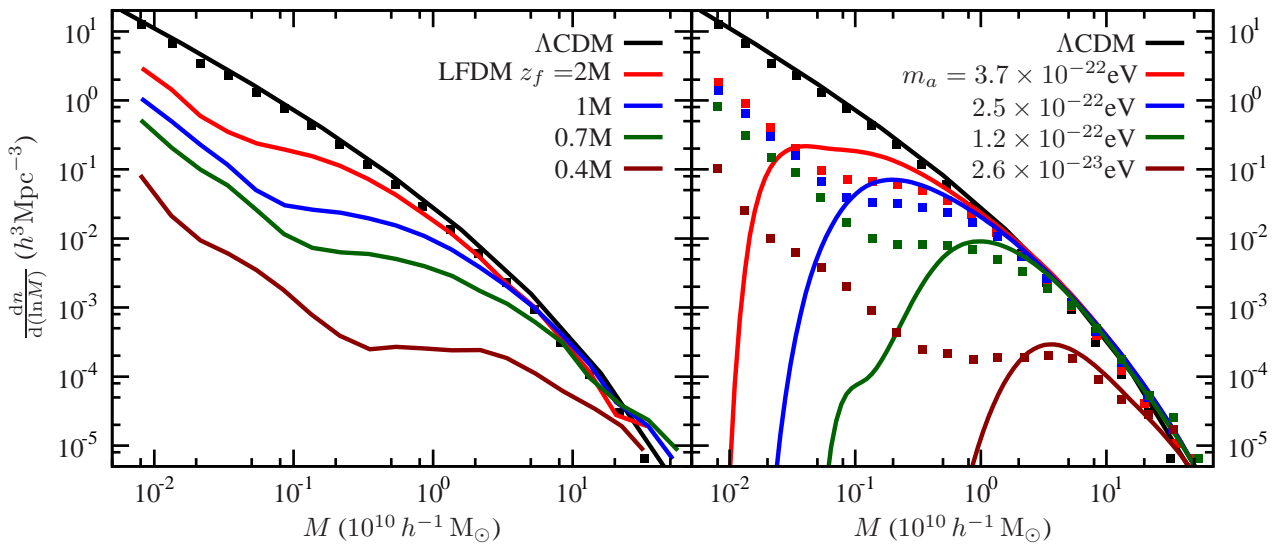}
\caption{This shows the halo mass function from our 
simulations considering four different LFDM models (left panel) with 
$z_f= \{2, 1, 0.7, 0.4\} \times 10^6$ and four different axion DM models 
(right panel) with axion masses $m_a=\{3.7, 2.5, 1.2, 0.26\} \times10^{-22}$ 
eV. The black solid curve is the theoretical $\Lambda$CDM 
mass function of Ref.~\cite{Sheth:2001dp}. In the right hand panel, the 
points represent the results of the simulation, while 
the solid curves are the theoretical prediction from  \citep{Marsh:2013ywa}.}
\label{fig:mass_func}
\end{figure}

Figures~\ref{fig:mass_func} shows the simulated comoving number density 
of halos per unit logarithmic halo mass ${\rm d}n/{\rm d(ln}M)$ as 
a function of the  halo mass $M$ at $z=8$ for the range of LFDM 
(left panel) and ULA (right panel) models that 
we consider here. The solid curve is the theoretical $\Lambda$CDM 
mass function \cite{Sheth:2001dp}. The simulated $\Lambda$CDM mass 
function, shown in black points, is in very good agreement with
the theoretical mass function. Note that the low mass halo abundance 
is substantially reduced for both the LFDM and the ULA models as 
compared to the $\Lambda$CDM, and the suppression gets steeper 
with decreasing value of $z_f$  and $m_a$ in the LFDM  and ULA models 
respectively. 
 
In the right panel of Figure~\ref{fig:mass_func} , we also show the theoretical halo mass function for ULAs of Ref.~\cite{Marsh:2013ywa}. 
The theoretical mass function displays a sharp cut-off at low halo masses caused by scale dependent growth and an increased barrier
for collapse: consequences of the ULA Jeans scale. This cut-off is not present in the mass function found from our N-body
simulations, but we should not expect it to be. N-body simulations treat ULAs as particles. However, this treatment is incomplete on small
scales, where the coherent scalar field dynamics become important. The full scalar field dynamics have been computed by Ref.~\cite{Schive:2014dra}, and the validity of an N-body treatment on large scales was discussed in Ref.~\cite{Schive:2015kza}.
The cut-off in the theoretical mass function for ULAs becomes relevant precisely where scale dependent growth and scalar field dynamics become important in linear theory. This suggests that the ULA mass functions we have derived should not be considered correct on these scales. 
 
Both the LFDM and ULA mass functions produce low mass halos below the cut-off in the linear theory power spectrum.
The same effect is observed in related simulations of warm dark matter (WDM). In the case of WDM, these low mass halos are
believed, for a variety of reasons, to be ``spurious"~\cite{ Wang:2007he, Lovell:2013ola}. We note that for ULAs the mass function slope from simulations increases below the cut-off in the theoretical mass function. An increase in the slope of the mass function is one method of identifying spurious halos~\cite{Schneider:2013ria}, and it is interesting that these scales coincide.
A complete simulation of ULAs (either as a scalar field~\cite{Schive:2014dra}, or an effective fluid \cite{Mocz:2015sda,Marsh:2015wka}) should include a dynamical mechanism whereby the spurious halos never form, thanks to the so-called ``quantum pressure" of the gradient energy. This further suggests, on theoretical grounds, that the low mass halos are spurious.  

Spurious structure can be removed from simulated mass functions: for WDM in e.g. Ref.~{\cite{Schneider:2013ria}}, for LFDM in Ref.~{\cite{Agarwal:2015iva}}, and for ULAs in Ref.~{\cite{Schive:2015kza}}. In our results we do not remove the spurious structure. The constraints thus derived will be weaker than the true constraints, allowing for lighter ULAs and later formation of DM. Our constraints are therefore, in some sense, conservative.

\subsection{Generating the neutral hydrogen (\HI) maps}
In the final step we generate the ionization map and 
the \HI distribution using the homogeneous 
recombination scheme of Ref~\cite{Choudhury:2008aw}.
The basic assumption here is that the hydrogen exactly traces 
the  matter density field and the  halos host the  
sources of ionizing photons. It is also assumed that the number 
of ionizing photons emitted by a source $N_\gamma$  is proportional to the 
mass of the host halo {\it i.e.} $N_\gamma = N_{\rm ion} M/m_p$ where $m_p$ is
the proton mass.  The constant of proportionality $ N_{\rm ion}$ 
here is  the number of  ionizing photons emitted 
per baryon in the collapsed halo 
times the ratio of the baryon density to  the total  matter density. 
The ionization map is generated by comparing the smoothed photon number density
to the smoothed  hydrogen number density at each grid point in the simulation
volume. Any grid point where the photon number density exceeds the hydrogen
number density is declared to be ionized. This comparison is carried out
varying the smoothing radius from the cell size to a maximum smoothing
length-scale which is half the box size.   The ionized map and the \HI
distribution were generated using a grid spacing that is $16$ times coarser
than the $N$-body simulations. The simulated \HI distribution was finally 
mapped to redshift space ({\cite{Bharadwaj:2004nr})  using the scheme outlined in 
Ref.~\cite{Majumdar:2012pk}. We use the resulting \HI distribution to calculate
the brightness temperature fluctuation using (\cite{Bharadwaj:2004it}) 
\begin{equation}
\delta T_b = 4{\rm mK}\, \,\frac{\rho_{\rm HI}}{\bar{\rho}_{\rm H}} (1+z)^2 \left(\frac{\Omega_b\,h^2}{0.02}\right)\left(\frac{0.7}{h}\right) \frac{H_0}{H(z)} \, , 
\label{eq:Tb}
\end{equation}
where $\frac{\rho_{\rm HI}}{\bar{\rho}_{\rm H}}$ is 
the ratio of the neutral hydrogen to the mean hydrogen 
density. Throughout this paper, we assume the spin temperature 
$T_s \gg T_{\rm cmbr}$ or the \HI is only observed in 
emission.

\section{Results}
\label{sec:result}
\begin{figure}
\centering
\includegraphics[width=1.05\textwidth, angle=0]{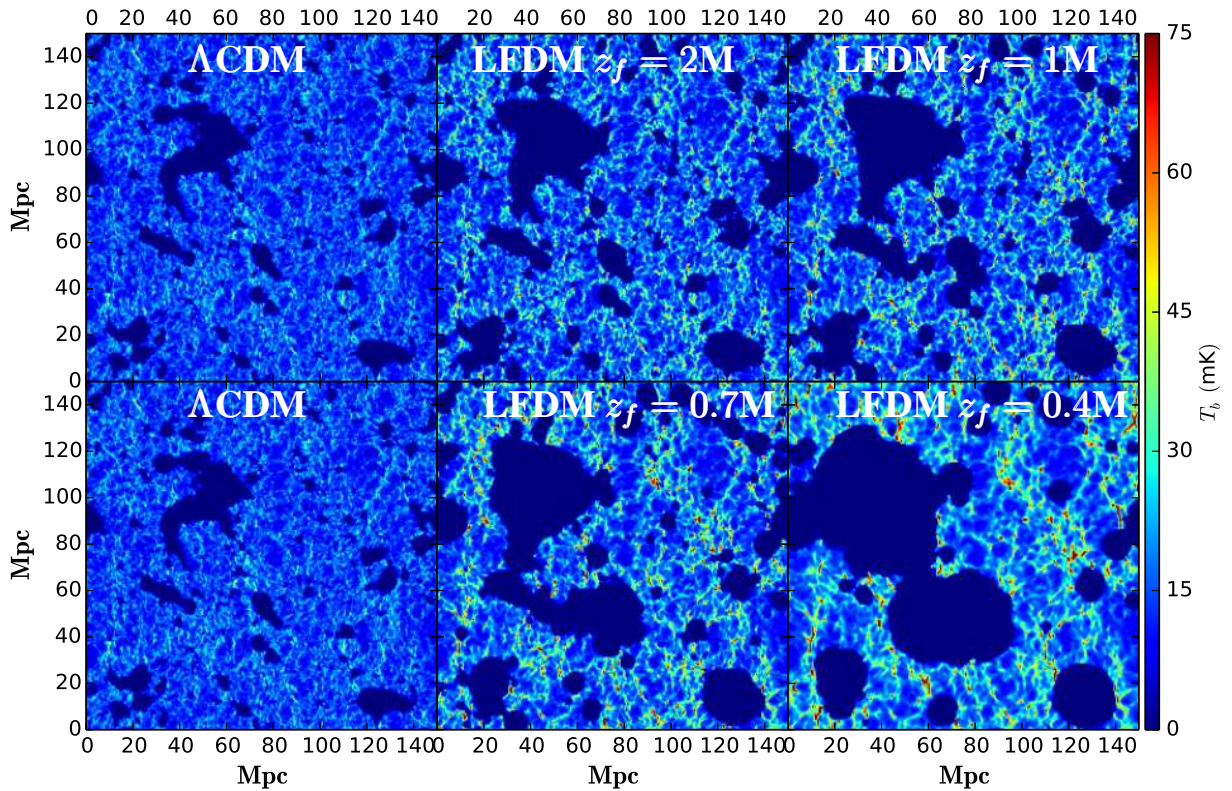}
\put(-391, 275){\textcolor{white}{\textbf{{\Large $\pmb{\Lambda}$CDM}}}}
\put(-285, 275){\textcolor{white}{\textbf{{\Large LFDM $\pmb{z_f=2}$M}}}}
\put(-149, 275){\textcolor{white}{\textbf{{\Large LFDM $\pmb{z_f=1}$M}}}}
\put(-391, 145){\textcolor{white}{\textbf{{\Large $\pmb{\Lambda}$CDM}}}}
\put(-281, 145){\textcolor{white}{\textbf{{\Large LFDM $\pmb{z_f=0.7}$M}}}}
\put(-149, 145){\textcolor{white}{\textbf{{\Large LFDM $\pmb{z_f=0.4}$M}}}}
\caption{Two dimensional sections through the simulated 
brightness temperature maps of four LFDM models along the 
$\Lambda$CDM model 
for $\xb=0.5$. The $\Lambda$CDM (left) map has been shown 
twice. The direction of redshift space distortion is with 
respect to a distant observer located along the vertical 
axis.}
\label{fig:h1_map}
\end{figure}
\begin{figure}
\centering
\includegraphics[width=1.05\textwidth, angle=0]{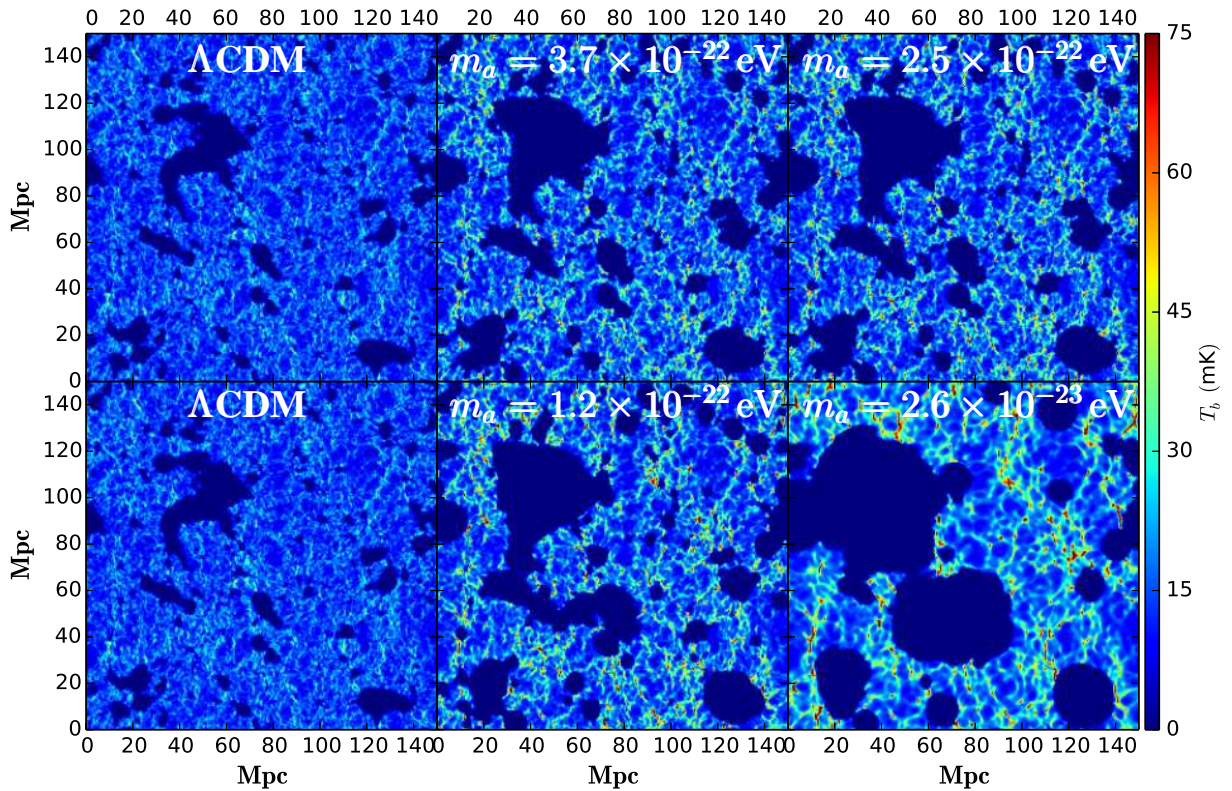}
\put(-391, 275){\textcolor{white}{\textbf{{\Large $\pmb{\Lambda}$CDM}}}}
\put(-293, 275){\textcolor{white}{\textbf{{\Large $\pmb{m_a=3.7 \times 10^{-22}\,{\rm eV}}$}}}}
\put(-161, 275){\textcolor{white}{\textbf{{\Large $\pmb{m_a=2.5 \times 10^{-22}\,{\rm eV}}$}}}}
\put(-391, 145){\textcolor{white}{\textbf{{\Large $\pmb{\Lambda}$CDM}}}}
\put(-293, 145){\textcolor{white}{\textbf{{\Large $\pmb{m_a=1.2 \times 10^{-22}\,{\rm eV}}$}}}}
\put(-161, 145){\textcolor{white}{\textbf{{\Large $\pmb{m_a=2.6 \times 10^{-23}\,{\rm eV}}$}}}}
\caption{Two dimensional sections through the simulated 
brightness temperature maps of the four different axion 
DM models with axion masses $m_a=3.7 \times 10^{-22}\,{\rm eV},\,
2.5 \times 10^{-22}{\rm eV},\,1.2 \times 10^{-22}{\rm eV}$ and $2.6 \times 10^{-23}{\rm eV}$ 
for $\xb=0.5$. The $\Lambda$CDM (left) map has been shown 
twice. The direction of redshift space distortion is with 
respect to a distant observer located along the vertical 
axis.}
\label{fig:h1_map_ULA}
\end{figure}
Our method predicts an `inside-out' ionization where the high 
density regions are ionized first and the low density region 
later. As the value of $z_f$ in LFDM model decreases, the 
number of halos decreases. This means that the number of 
ionizing sources in the LFDM and ULA  models is smaller  as
 compared  to $\Lambda$CDM 
model. To achieve the same ionization level at the same 
redshift   ($\xb=0.5$), $N_{\rm ion}$ is higher 
in the LFDM models as compared to the $\Lambda$CDM model. 
We find 
that for $z_f < 0.4 \times 10^6$ the desired level of reionization doesn't 
occur as we couldn't form dark  matter halos in our box. This allows us to  put 
a rough limit of $z_f \sim 0.4 \times 10^6$ as  a 
lower cut off. Similar considerations allows us to put a lower limit
on the mass of ULA of $m_a > 2.6 \times  10^{-23} \, \rm eV$. It is instructive to note 
that this result is consistent with \cite{Hlozek:2014lca,Schive:2015kza} 
in connection to linear as well as non-linear observables of ULA DM. Table \ref{table:1} 
lists the set of models  we study in this paper. 

However, we could also get independent limit on $z_f$  and $m_a$
from constraints on plausible  range of $N_{\rm ion}$. In our simulation 
$N_{\rm ion} = 12$
for the $\Lambda$CDM model. For the range of LFDM models
we have studied (for decreasing $z_f$ as shown in Figure~\ref{fig:h1_map}): 
 $N_{\rm ion} = \{37, 95, 207, 1401\}$. For ULA models,  for decreasing $m_a$ as 
shown in Figure~\ref{fig:h1_map_ULA}, $N_{\rm ion} = \{53, 70, 130, 1140\}$. 
Table \ref{table:1} lists the values of $N_{\rm ion}$ for the models we consider. 

%%%%%%%%%%%%%%%%%%%%%%%%%%%%%%%%%%%%%%%%% 
\begin{table}
\begin{center}
 \begin{tabular}{|c|c|c|c|} 
 \hline
 Model     & Parameter     & $N_{\rm ion}$     & Reionization \\ [1.5ex] 
 \hline
 CDM     & \cite{Ade:2015xua,Ade:2013zuv}     & 12     &     $\surd$ \\ [1.5ex]
 \hline
 \,     & $z_f=2.0$M & 37       & $\surd$     \\ [1.5ex] \cline{2-4}
 \,     & $z_f=1.0$M & 95      & $\surd$     \\ [1.5ex] \cline{2-4}
 LFDM   & $z_f=0.7$M & 207      & $\surd$     \\ [1.5ex] \cline{2-4}
 \,		& $z_f=0.4$M & 1401     & $\times$     \\ [1.5ex] \cline{2-4}
 \,		& $z_f=0.2$M &  No Haloes     & $\times$    \\ [1.5ex]
 \hline
 \,     & $m_a=3.7 \times 10^{-22}$eV     & 53     & $\surd$   \\ [1.5ex] \cline{2-4}
 \,     & $m_a=2.5 \times 10^{-22}$eV     & 70     & $\surd$   \\ [1.5ex] \cline{2-4}
 ULA DM & $m_a=1.2 \times 10^{-22}$eV     & 130    & $\surd$   \\ [1.5ex] \cline{2-4}
 \,     & $m_a=2.6 \times 10^{-23}$eV     & 1140    & $\times$   \\ [1.5ex] \cline{2-4}
 \,     & $m_a=2.0 \times 10^{-23}$eV     & No Haloes    & $\times$  \\ [1.5ex]
 \hline
\end{tabular}
\end{center}
\caption{The Table lists the values of $N_{\rm ion}$ for the LFDM and ULA models
we consider. In the last column the tick mark illustrates  whether the models is able to achieve reionization  based on 
 an acceptable value of $N_{\rm ion}$ and the formation of haloes in the N-body simulation.}
\label{table:1}
\end{table}
%%%%%%%%%%%%%%%%%%%%%%%%%%%%%%%%%%%%%%%%%%
 
How acceptable are these values if star-forming galaxies were responsible for
the reionization process? For a metallicity $Z = 0.01$ and Scalo stellar mass
function, the number of hydrogen ionizing photons is nearly 4000 per baryon
which corresponds to $N_{\rm ion} \simeq 800$. Factoring in $10\%$ star formation
efficiency in a halo for star forming galaxies 
 and $10\%$ escape fraction from these haloes, this 
number drops by a factor of 100. \footnote{We note that the effective number of 
hydrogen ionizing photons in the case of early QSOs where these photons 
are produced owing to the conversion of gravitational energy into energy
are comparable to the case of early star-forming galaxies \cite{Barkana:2000fd}.}  It should be noted that all these factors---
metallicity, initial mass function, star formation efficiency, and escape fraction---are highly uncertain. For zero metallicity (population III) stars the 
number of photons could be significantly higher  and lie in the 
range  $10^4\hbox{--}10^5$ \footnote{From current observations it is 
difficult to constrain the fraction of PopIII stars during the 
epoch of reionization but plausible bounds based on the observed Infra-red
background and its fluctuations suggest PopIII stars  might not have dominated 
the reionization process \cite{Salvaterra:2005ga, Yue:2012dd}}, but these stars last only a 
few million years which is considerably smaller than the age of the universe, 
$\simeq 5 \times 10^8 \, \rm yrs$ at $z \simeq 8$. 
These   first metal-free, massive stars would have  ended their   life in
supernova explosions thereby  
contaminating  the interstellar medium with metals. This means the metal-free stars could only have dominated  the reionization process for   short periods.  Also 
for initial stellar mass functions which have a larger fraction  of 
 massive O and B stars as compared 
to the Scalo mass function, the number of ionizing photons  could be larger
(for details of the physics of ionizing sources during the 
epoch of reionization  see \cite{Barkana:2000fd}). 

In light of these facts we could ask how plausible is $N_{\rm ion}$ corresponding
to $z_f \simeq 0.4 \times 10^6$ or $m_a \simeq 2.6 \times 10^{-23} \, \rm eV$. In these cases, the required number of photons per 
baryon is larger than 5000. This is not possible to achieve  for moderate metallicities 
and Scalo mass function even if the efficiency of star formation and 
the escape fraction are 100\%. Therefore, we consider such models unrealistic.
 
Upcoming near-infrared telescope JWST will allow us to directly detect
ionizing sources from the epoch of reionization (e.g. Figure 19 of \cite{Barkana:2000fd}). For LFDM and ULA models, these sources are fewer in number and 
more luminous, which might allow a direct probe of the decrement of matter  power  spectra at small scales. 

Figures~\ref{fig:h1_map} and~\ref{fig:h1_map_ULA} show two-dimensional 
 sections through the simulated 
brightness temperature cubes  for  LFDM and ULA  models 
 for $\xb=0.5$. By visual 
comparison we see two main differences between $\Lambda$CDM 
and LFDM (ULA) models. The first difference 
is that the size of the ionized regions is larger in the LFDM (ULA)
models. It is owing to two factors: 
first, as discussed above, it  is a consequence of the fact that the sources 
require higher star formation  
efficiency to achieve the desired level of ionization. Second, the suppression of matter power at small scales
results in a decrement of mass dispersion at these scales. Therefore, the 
haloes that form are a higher $\sigma$ fluctuations of the density field  as compared to the 
$\Lambda$CDM model. It is known that higher $\sigma$ fluctuations are 
more strongly clustered for a Gaussian field (see e.g. \cite{Kaiser:1984sw,peebles:1993}), or 
the ionizing sources are formed more  preferentially in a cluster. 
Both these factors contribute to enlarging the size of the ionizing bubble
and explain why the  ionized bubble sizes  get  larger 
 with decreasing value of $z_f$  ($m_a$) in the LFDM (ULA)  models. 
The second difference, also linked to the factors discussed above, 
 is that the \HI\, fields has stronger
 contrast in the LFDM models. 
Both these differences manifest themselves in the power spectra of the HI 
field which we discuss next. 
\begin{figure}
\psfrag{pk}[c][c][1][0]{$\Delta_{\rm b}^2\, (k)$}
\psfrag{k}[c][c][1][0]{$k\,$ (${\rm Mpc}^{-1}$)}
\psfrag{10}[c][c][1][0]{10}
\psfrag{CDM}[c][c][1][0]{{$\Lambda$CDM}\, \,}
\psfrag{LFDM z=2M}[c][c][1][0]{{LFDM $z_f=$2M}\, \, \, \,}
\psfrag{1M}[c][c][1][0]{{1M}}
\psfrag{0.7M}[c][c][1][0]{{0.7M}}
\psfrag{0.4M}[c][c][1][0]{{0.4M}}
\psfrag{ULA mw14}[c][c][1][0]{{ULA $m_a=3.7\times10^{-22}$eV}\, \, \, \, \, \, \, \, \, \, \, \, \, \, \, \,}
\psfrag{mw12}[c][c][1][0]{{$2.5\times10^{-22}$eV}\, \, \, \, \, \, \, \, \,}
\psfrag{mw09}[c][c][1][0]{{$1.2\times10^{-22}$eV}\, \, \, \, \, \, \, \, \,}
\psfrag{mw05}[c][c][1][0]{{$2.6\times10^{-23}$eV}\, \, \, \, \, \, \, \, \,}
\centering
\includegraphics[width=1.05\textwidth]{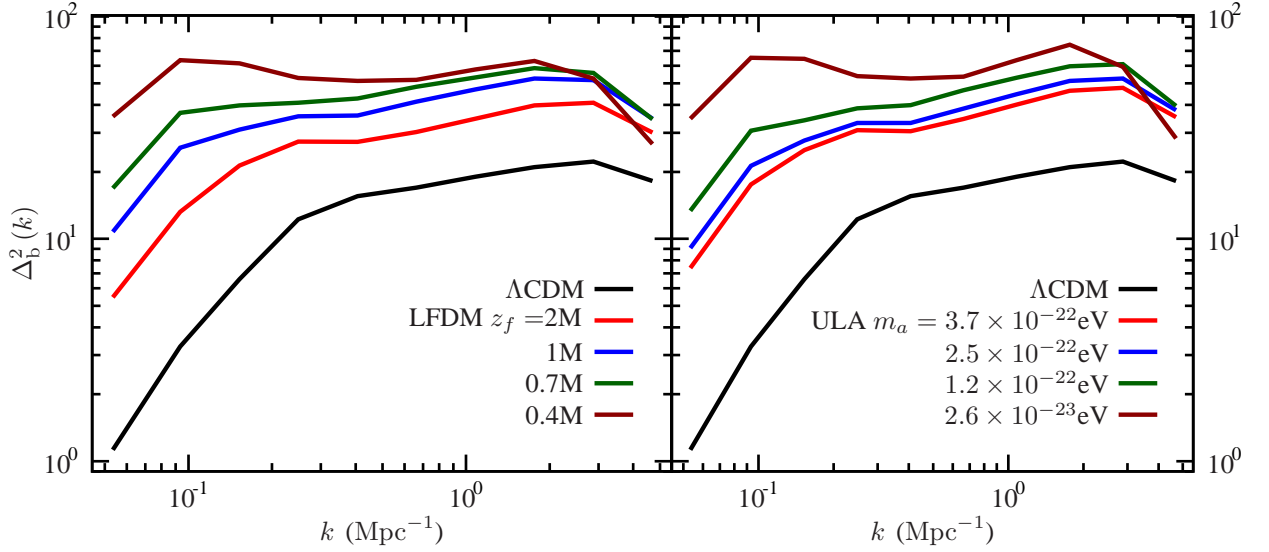}
\caption{This shows the brightness temperature power 
spectrum $\Delta_{\rm b}^2\, (k)$ ($\rm mK^2$) of the \HI\, field 
for the four different LFDM models (left panel) with 
$z_f= \{2, 1, 0.7, 0.4\} \times 10^6$ and four different axion DM models 
(right panel) with axion masses $m_a=\{3.7, 2.5, 1.2, 0.26\} \times10^{-22}$ 
eV. The black solid curves are for the $\Lambda$CDM model.}
\label{fig:pk_ionz}
\end{figure}

Figures~\ref{fig:pk_ionz} shows the mean squared brightness 
temperature fluctuation $\Delta_{\rm b}^2\, (k)=k^3 P_{\rm b}
\,(k)/2 {\rm \pi}^2$ of the \HI field for LFDM (left panel) 
and ULA (right panel) models
we consider here along with the  $\Lambda$CDM  model, 
for a fixed ionization fraction  $\xb=0.5$. We find that the power for LFDM (ULA) models is greater than 
the $\Lambda$CDM model over a large range of scales  $0.1<k<4 \, \rm Mpc^{-1}$. 
This is owing to the factors discussed in the foregoing. 
The scale of the HI power spectra from ionization inhomogeneities, 
 is governed by the 
size of  ionized bubbles. These 
inhomogeneities normally dominate the contribution of density perturbations
for scales considered here. Therefore   we expect, for 
the same ionized fraction, 
the power to increase if the ionized bubbles are larger. It has been 
shown analytically \cite{Furlanetto:2004nh} and is consistent with the results of 
numerical simulations  \cite{Lidz:2007az} (see also \cite{Shao:2012cg}). Conversely, when there is 
enhancement of matter power at small scales, e.g.  owing to the presence of 
primordial magnetic fields, the HI  signal in the range of scale 
discussed here, diminishes \cite{Sethi:2009dd}. 

 We only consider a fixed ionization fraction $\xb =0.5$ at $z = 8$ for 
our study. However, the discussion in the foregoing shows  that the enhancement 
of the HI signal is generic and should be independent of the ionization 
fraction. It can be shown that the HI signal is dominated by ionization
inhomogeneities and it peaks at close to $\xb = 0.5$ \cite{Furlanetto:2004nh}, 
which also  partly motivates our choice.

Current observational constraints from radio interferometer PAPER put an
upper limit on the HI bright temperature power spectrum: $\Delta_b^2(k) < (22.4 {\rm mK})^2$ over the range of scales  $0.15 < k < 0.5 \rm k Mpc^{-1}$ at $z = 8.4$ \cite{Pober:2015ema}.  These constraints are too weak, by roughly a factor of 10, to probe the enhancement of the HI signal
seen in Figure~\ref{fig:pk_ionz} for LFDM and ULA models at the present.
Ongoing radio interferometers focused on detecting the HI signal
from EoR---LOFAR, MWA, and PAPER---might throw more light on this issue 
in the near future.

\section{Evolution of cosmological gas density}
\label{sec:LFDMonCosmo}

As discussed in the previous section, the main impact of LFDM and ULA models
is to reduce the matter  power at small scales. This 
results in  smaller number of haloes
at masses relevant for studying the reionization process. This also means
the collapsed fraction of matter decreases for models with lower 
power. We study the implication of the 
evolution of the collapsed fraction for LFDM and Axion models in this section.

From absorption studies  of Damped Lyman-$\alpha$ clouds, the evolution 
of average mass density of  HI in the universe can be inferred \cite{Noterdaeme:2009tp,Peroux:2001ca}.  Assuming 
that the HI follows baryons and the  collapsed fraction of baryons
traces dark matter, this allows us to get an approximate measure of the 
minimum amount of collapsed fraction of the total matter in the redshift
range $2 < z < 5$ for which the data are  available \cite{Noterdaeme:2012gi,Noterdaeme:2009tp,Zafar:2013bha}. From the HI data one obtains, $\Omega_{\rm HI}(z) = \rho_{\rm HI}^{\rm coll}(z)/\rho_c(z)$, which gives the fraction of the collapsed neutral
hydrogen in terms of critical density of the universe $\rho_c$. The (minimum)  collapsed fraction is given by: $f_{\rm coll}(z) = \rho_c(z)/\rho_b(z) \Omega_{\rm HI}(z)$, where $\rho_b$ is the background energy density of  Baryons. 

As obtaining  the mass function from N-body simulation is numerically expensive, for computing the collapsed fraction, for LFDM models, we integrate 
 the Sheth-Tormen mass function \cite{Sheth:1999mn,Sheth:2001dp} above the density threshold of collapse at a given redshift. The collapsed fraction (the fraction of collapsed mass  in haloes  with masses larger than  $M$) at a redshift
$z$ is given by: 
\begin{equation}
f_{\rm coll}(M,z) = \int_{\nu_{\rm min}}^\infty \nu f(\nu) {d\nu \over \nu}
\label{eq:collfra}
\end{equation}
Here $\nu_{\rm min} = (1.69/\sigma(M,z))^2$ and $f(\nu)$ is given by the 
Sheth-Tormen mass function \cite{Sheth:1999mn,Sheth:2001dp}. For computing the collapsed fraction for 
ULA models, we integrate the halo mass functions derived by \citep{Marsh:2013ywa}. 

In Figures~\ref{fig:3} and~\ref{fig:3ax} we show the mass dispersion $\sigma(M)$  for a range of LFDM and ULA models
and collapsed fraction as a function of mass for a fixed redshift. It is 
seen that as the formation redshift $z_f$  ($m_a$) decreases, $\sigma(M)$ decreases for masses which correspond
to scales at which there is decrement in power. This also means that collapse
fraction falls as $z_f$ ($m_a$)  decreases. 

It is not straightforward to compare the theoretical  collapsed fraction (Eq.~(\ref{eq:collfra})  with the damped
Lyman-$\alpha$ data because there is a large uncertainty in the masses of 
these clouds. Simulations suggest that the mass of damped-$\alpha$ clouds
could lie in the range $10^9\hbox{--}10^{10} \, \rm M_\odot$ (\cite{Pontzen:2008mx}).
However, recent observations suggest that the mass could be as high as 
$10^{12} \, \rm M_\odot$ at $z \simeq 2.5$. (\cite{FontRibera:2012fm}).  For the present 
work, we assume two halo masses $10^{10}$ and $5\times 10^{10} \, \rm M_\odot$ as the threshold
masses for the formation of Damped Lyman-$\alpha$ clouds. We compute the collapsed fraction for comparison with the  data by 
integrating the mass function with  threshold mass  as lower limits. 

In Figures~\ref{fig:4} and~\ref{fig:4ax} we compare the prediction of LFDM  and ULA models with the damped Lyman-$\alpha$ data. As noted above, the damped Lyman-$\alpha$ data provide a lower limit to the 
collapsed fraction. Therefore, all models that predict collapsed  fraction smaller
than the data can be ruled out. As expected, the constraints are tighter for the higher threshold 
halo mass. Figures~\ref{fig:4} and~\ref{fig:4ax}  show  that  $z_f < 2 \times 10^5$ and $m_a < 10^{-23} \, \rm eV$ can be ruled out
from the present data. 
 It should be noted that the data at higher redshifts provide tighter bounds.  

In the foregoing we discussed  the impact modified dark matter 
models  on the collapsed fraction of HI 
at high redshifts. It is possible to observe the 
collapsed fraction of matter   at high redshifts in other wave-bands 
in  the form of  luminosity function of  galaxies. Such 
observations  have been considered to constrain the ULA models \cite{Bozek:2014uqa}. Here we briefly discuss the   possible constraints on dark matter 
models  
from the currently existing   observations of the luminosity function of high redshift
galaxies. 
  
Many groups have recently  determined the luminosity function of bright galaxies,  using colour-based Lyman-break selection criterion  to determine the redshift, in  the redshift range $z = 7\hbox{--}8$ \cite{Bouwens:2010gp,Trenti:2010sz,Bouwens:2014fua}. We focus on the paper of Bouwens et al. (2015) \cite{Bouwens:2014fua} 
 for our discussion. Their determination of luminosity functions of galaxies is based 
on 481 sources at $z\simeq7$ and 217 sources at $z \simeq 8$. For $z \simeq 7$ they cover the absolute magnitude range
$M_{AB} = -22.16\hbox{--}-16.91$ and  the range for $z \simeq 8$  is  $M_{AB} = -21.87\hbox{--}-17.62$. To constrain dark matter 
models we need to find a relation between the luminosity function and the mass function of haloes 
predicted by these models. It should be underlined that the mass of haloes
is not measured directly in such observations and therefore such a relation
is based on empirical fitting based on mass-luminosity relations at lower
redshifts.  In Appendix I of Bouwens et~al. (2015)   paper \cite{Bouwens:2014fua}, they propose a redshift-dependent kernel to 
relate the measured luminosity functions with mass functions. They propose a relation between  mass 
and  luminosity based on a log-normal kernel and claim an acceptable fit in the redshift range $z = 4\hbox{--}8$ and reasonable agreement with earlier results. Their Eq.~I2 gives a  redshift-dependent  relation that can be used to approximately determine mass given the luminosity (this relation is not unique but the fluctuation 
in this relation is small,   Eq. I1 of Bouwens et al. (2015)). For the range of absolute magnitudes for which the luminosity functions 
have been determined this gives  the mass range: $M \simeq 2\times 10^{10}\hbox{--}10^{12} \, \rm M_\odot$ at $z \simeq 7$ and $M = 3\times10^{10}\hbox{--}10^{12} \, \rm  M_\odot$ at $z \simeq 8$.

We can compare this mass range with our results in Figures~7 and~8. These figures show that for most of the 
models (both LFDM and ULA) we consider, the predictions for the  mass range of interest are in reasonable agreement with the LCDM model, even though they differ significantly for smaller masses. For the mass range suggested by
measured luminosity functions for $z \simeq 7 \hbox{--}8$,  there is substantial
difference between the LCDM model and the modified dark matter for  
$m_a \simeq  2.6 \times 10^{-23} \, \rm eV$ and $z_f < 2 \times 10^5$. This is suggestive that 
these models are ruled out by high-redshift luminosity function measurements. These constraints are in reasonable agreement with those obtained 
from  other observables we consider in this paper.

\begin{figure}[H]
\begin{subfigure}{.5\textwidth}
  \centering
  \includegraphics[width=0.75 \linewidth , angle=-90]{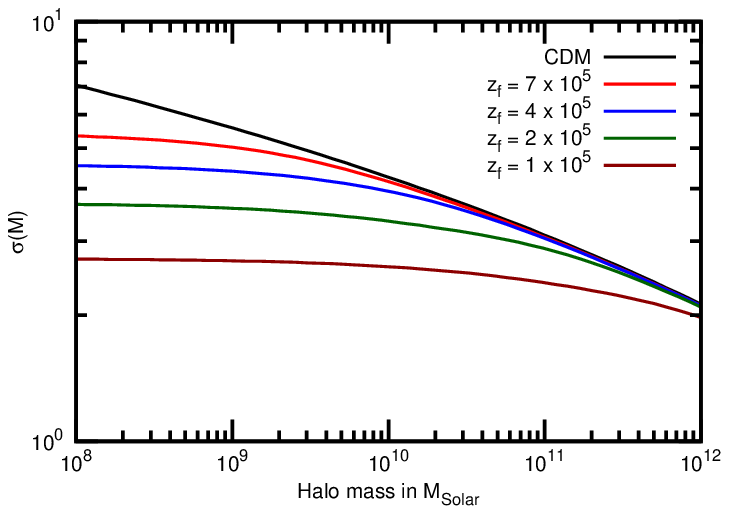}
  \label{fig:3a}
\end{subfigure}%
\begin{subfigure}{.5\textwidth}
  \centering
  \includegraphics[width=0.75 \linewidth , angle=-90]{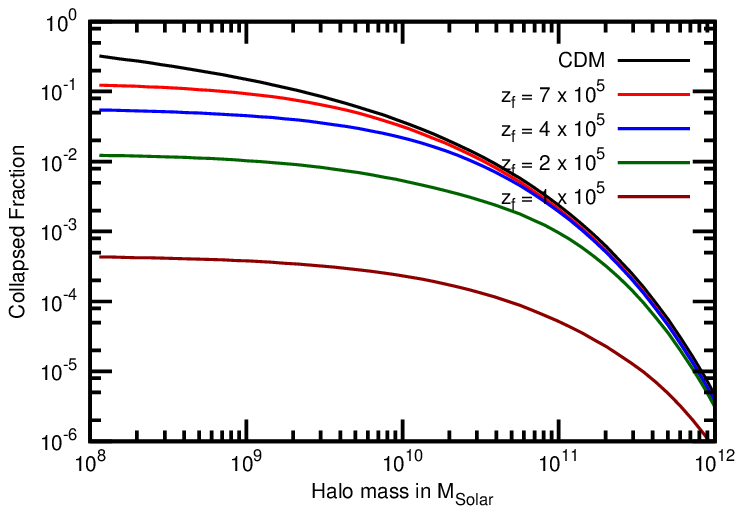}
  \label{fig:3b}
\end{subfigure}
\caption{The mass dispersion  $\sigma_M$ is shown for 
a range of LFDM models (Left Panel). The right
panel shows the collapse fraction  as a function of halo mass.  
Both the panels are for $z = 6$.}
\label{fig:3}
\end{figure}

\begin{figure}[H]
\begin{subfigure}{.5\textwidth}
  \centering
  \includegraphics[width=0.75 \linewidth , angle=-90]{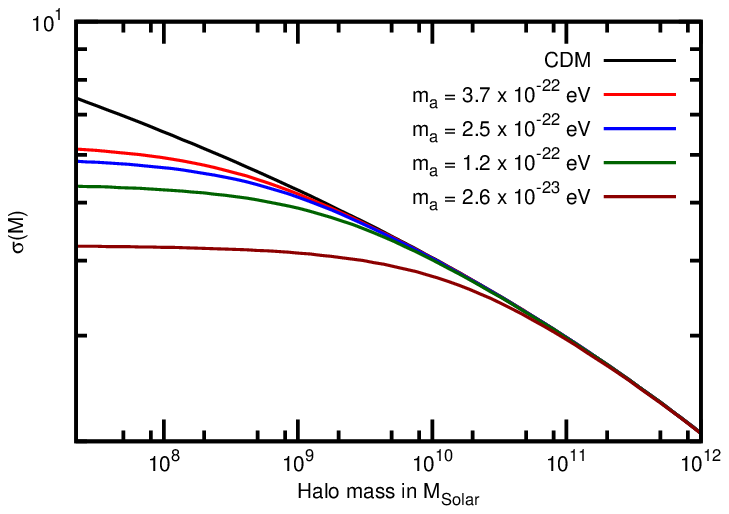}
  \label{fig:3aa}
\end{subfigure}%
\begin{subfigure}{.5\textwidth}
  \centering
  \includegraphics[width=0.75 \linewidth , angle=-90]{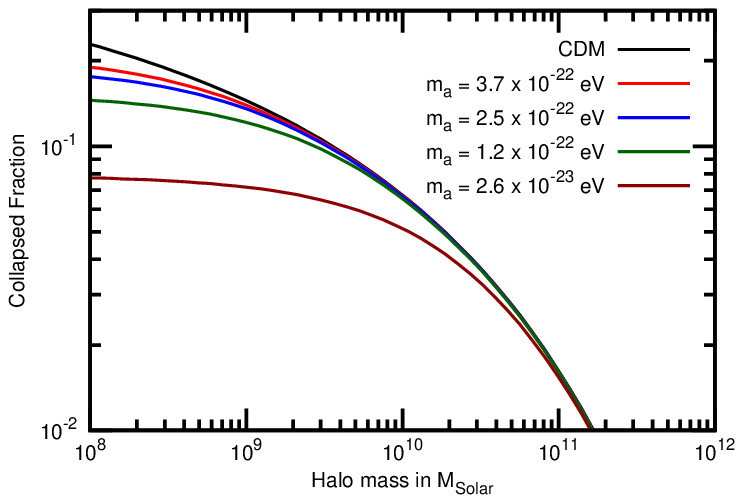}
  \label{fig:3ab}
\end{subfigure}
\caption{The same as Figure~\ref{fig:3} for ULA models.}
\label{fig:3ax}
\end{figure}
\begin{figure}[H]
\begin{subfigure}{.5\textwidth}
  \centering
  \includegraphics[width=0.70\linewidth , angle=-90]{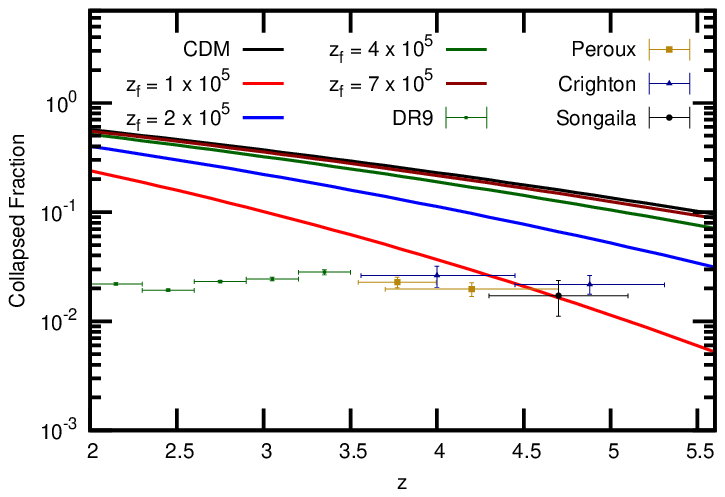}
  \label{fig:4a}
\end{subfigure}%
\begin{subfigure}{.5\textwidth}
  \centering
  \includegraphics[width=0.70 \linewidth , angle=-90]{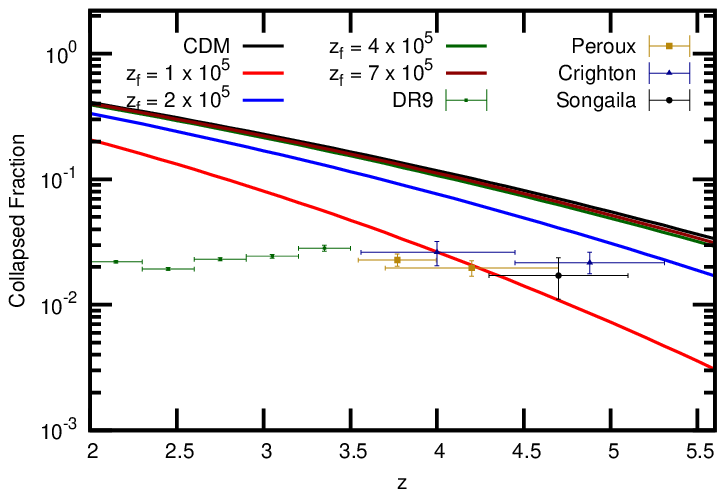}
  \label{fig:4b}
\end{subfigure}
\caption{The redshift evolution of collapsed fraction for LFDM models
is shown along with with the data  from Damped Lyman-$\alpha$ absorption data  \cite{Noterdaeme:2012gi,Noterdaeme:2009tp,Zafar:2013bha,Crighton:2015pza,Songaila:2010bu}. The left (right)  panels correspond to  threshold halo masses 
$M = 10^{10} \, \rm M_\odot$ ($M = 5 \times 10^{10} \, \rm M_\odot$), respectively.}
\label{fig:4}
\end{figure}

\begin{figure}[H]
\begin{subfigure}{.5\textwidth}
  \centering
  \includegraphics[width=0.70\linewidth , angle=-90]{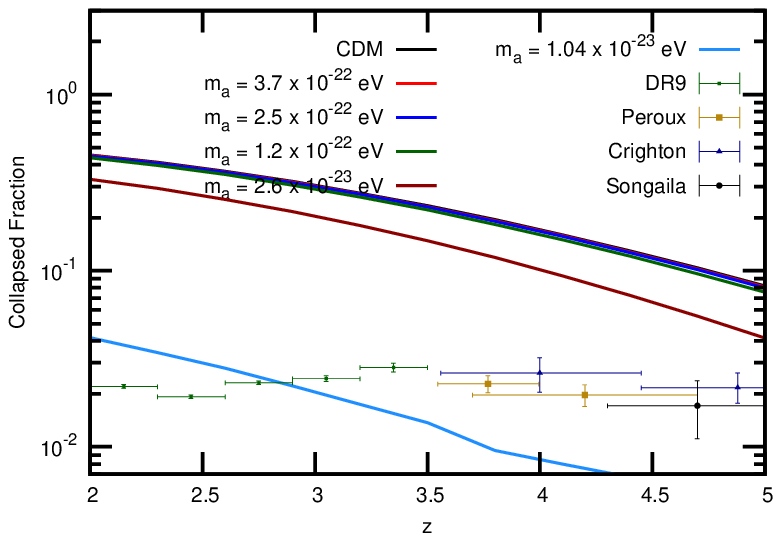}
  \label{fig:4aa}
\end{subfigure}%
\begin{subfigure}{.5\textwidth}
  \centering
  \includegraphics[width=0.70 \linewidth , angle=-90]{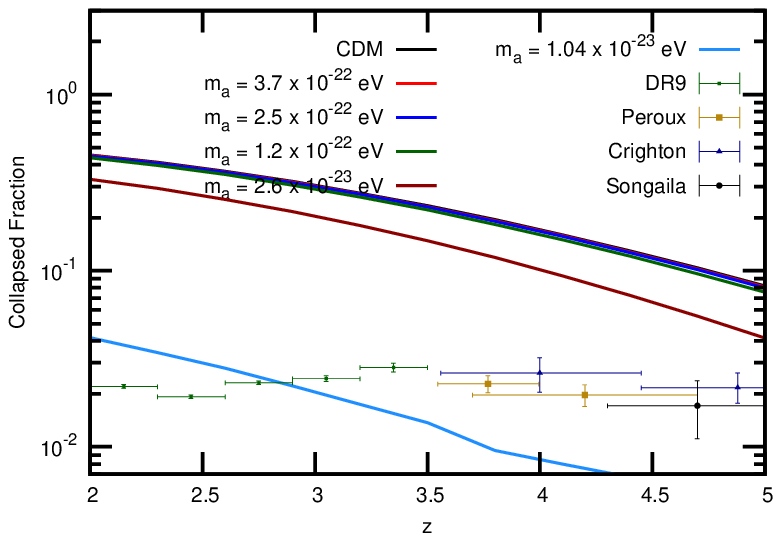}
  \label{fig:4ab}
\end{subfigure}
\caption{The same as Figure~\ref{fig:4} for ULA models}
\label{fig:4ax}
\end{figure}

\section{Conclusion}
\label{sec:conclude}
%
%

%
%N_ion = 47 for LFDM 2_ f =2M;   N_ion = 61 for ULA  m_a = 3.7 × 10 −22 eV
%N_ion = 102 for LFDM 2_ f =1M;   N_ion = 75 for ULA  m_a = 2.5 × 10 −22 eV
%N_ion = 212 for LFDM 2_ f =0.7M;   N_ion = 134 for ULA  m_a = 1.2 × 10 −22 eV
%N_ion = 1230 for LFDM 2_ f =0.4M;   N_ion = 965 for ULA  m_a = 0.26 × 10 −22 eV
%
From cosmological observations we know that DM must display clustering properties similar to CDM, at least on the largest scales.
 However, the particle nature of DM is still not
known. The linear LSS observations, such as WiggleZ, BOSS,  and Planck data on CMB temperature and polarization anisotropies  can be used to place very stringent constraints
on any dark matter models that show deviation from the $\Lambda$CDM model for  $k < 0.1 \, \rm Mpc^{-1}$ \cite{Hlozek:2014lca, Ade:2013zuv, Ade:2015xua}. However, at smaller scales the 
situation remains unclear from astrophysical observables, e.g. the missing satellite problem \cite{Klypin:1999uc,Moore:1999nt}. This points to a need to study
plausible dark matter models that allow for significant deviation from $\Lambda$CDM models at small scales and confront them with astrophysical 
and cosmological observables. This paper is one step in that direction.

In this paper, we consider the implications, and possible constraints on, of 
a class of  LFDM  and  axion-inspired models by running a series of N-body and reionization field numerical simulations with the appropriate power spectra as initial conditions.  Specifically, we study 
 the  reionization epoch  and the 
evolution of the collapsed fraction of matter at high redshifts. Such models generically 
give a decrement in matter power at small scales. In particular, we study 
the HI signal from the epoch of reionization.  We also consider the 
evolution of the collapsed fraction of the matter in the redshift range
$2 <z < 5$, using the data of damped Lyman-$\alpha$ on $\Omega_{\rm HI}(z)$.

We show that the power spectrum
of the HI field could be higher  for such models for a fixed ionization 
fraction as compared to the $\Lambda$CDM model. The enhancement is by 
factors of  $2\hbox{--}10$ for range of scales $0.1 < k < 4 \, \rm Mpc^{-1}$. (Figure~\ref{fig:pk_ionz}). 

For studying the EoR with alternative DM models, we demand that 
these models be  able to provide a reionization fraction $x=0.5$ at $z=8$. For very low $z_f$ and $m_a$, no halos  are  formed at $z=8$ in the N-body box, which rules out of these
models. In models with larger $z_f$ and $m_a$, the desired amount of reionization is achieved by increasing  the number of ionizing photons, $N_{\rm ion}$ as 
compared to the $\Lambda$CDM model. The models that  
 require  unrealistically large value of the number of 
ionizing photons   are also  excluded. These considerations are listed  in Table \ref{table:1}  and lead to  generic  bounds
on the epoch of the formation of dark matter $z_f > 4 \times 10^5$ and 
the axion mass $m_a > 2.6 \times 10^{-23} \, \rm eV$. We also obtain
weaker constraints from  the damped Lyman-$\alpha$ data on the 
evolution of collapsed gas 
fraction: $z_f >  2 \times 10^5$  and $m_a > 10^{-23} \rm \, eV$. We also argue
that the observed luminosity functions of high redshift galaxies in the redshift
range $z \simeq 7\hbox{--}8$ are also expected to yield similar constraints. 

The models we consider in this paper have been studied 
in the context of other cosmological observables. The LFDM models 
have been compared against the SDSS galaxy clustering and the Lyman-$\alpha$ data  \cite{Sarkar:2014bca}; the SDSS data gives $z_f > 10^5$ while the Lyman-$\alpha$ data results in stronger bounds $z_f > 9 \times 10^5$ (all 3$\sigma$). While the Lyman-$\alpha$ data give more stringent constraints than we obtain here,
the models consistent with this data still result in an enhancement of up to 
a factor of 4 in the HI power spectra (Figure~\ref{fig:pk_ionz}). The ULA models
considered here have been confronted with 
 Lyman-$\alpha$ and galaxy luminosity 
data; the resulting bounds are  $m_a > 10^{-22} \, \rm eV$ which  are  comparable to the constraints
we obtain in this paper \cite{Amendola:2005ad,Schive:2015kza}. It is interesting to note that the current constraints on ULAs,
including our own, are consistent with the mass necessary for a successful solution of the cusp-core problem in dwarf
spheroidals \cite{Walker:2011zu, Schive:2014hza, Schive:2014dra, Marsh:2015wka}.

Building a  physical model  of the nature of  dark matter   consistent with 
all the astrophysical and cosmological observables has not been achieved yet. 
Cosmological observables affected by  small scale matter power remain 
 crucial elements in this quest. In this paper we studied the impact of two
such observables for two classes of non-WIMP dark matter models and were able to constrain their underlying parameters. 
 In the future we hope to return to this issue with further comparisons with 
the low and high  redshift astrophysical  data. 

\section*{Acknowledgements}
The authors would like to thank the anonymous referee for the valuable constructive comments. DJEM acknowledges the support of a Royal Astronomical 
Society post-doctoral research fellowship, hosted at King's College London. DJEM also acknowledges useful discussions with Hsi-Yu Schive.

\bibliographystyle{JHEP}
\bibliography{paper_v3}{}

\end{document}